\documentclass[aps,prl,twocolumn]{revtex4}
\usepackage[dvips]{graphicx}

\begin{document}
\title{
Periodicity and criticality in the Olami-Feder-Christensen model of earthquakes
}

\author{
Takeshi Kotani, Hajime Yoshino and Hikaru Kawamura
}

\affiliation{Department of Earth and Space Science, Faculty of Science,
Osaka University, Toyonaka 560-0043,
Japan}
\date{\today}
\begin{abstract}
Characteristic versus critical features of earthquakes are studied on the basis of the Olami-Feder-Christensen model. It is found that the local recurrence-time distribution exhibits a sharp $\delta$-function-like peak corresponding to rhythmic recurrence of events with a fixed ``period'' uniquely determined by the transmission parameter of the model, together with a power-law-like tail corresponding to scale-free recurrence of events. 
The model exhibits phenomena closely resembling the asperity known in seismology.
\end{abstract}
\maketitle
 Earthquakes possess two apparently contrasting features. In one, earthquakes exhibit {\it critical\/} features. Earthquake occurrence is often characterized by power laws, {\it e.g.\/}, the Gutenberg-Richter (GR) law or the Omori law. As is well-known, phenomena described by power-laws are scale-invariant without any characteristic energy or time scale. In the other, earthquakes are sometimes regarded to possess {\it characteristic\/} features, with a characteristic energy or time scale. The relation between these apparently contrasting views of earthquakes, {\it i.e.\/}, critical versus characteristic, still remains to be understood \cite{Scholzbook,Kolkotta}.

  Modeling earthquakes and elucidating its statistical properties have been a fruitful strategy in earthquake studies. One of the standard model of earthquakes might be the so-called spring-block or the Burridge-Knopoff (BK) model, in which earthquake fault is modeled as an assembly of blocks mutually connected via elastic springs which are slowly driven by external force \cite{BK,Carlson}. Olami, Feder and Christensen (OFC) introduced a further simplified  model,  called the OFC model \cite{OFC}, mimicking the BK model. It is a two-dimensional lattice model where the rupture propagates from lattice site to its nearest-neighboring sites  in a non-conservative manner, often causing multi-site ``avalanches''. Extensive numerical studies have revealed that the OFC model exhibits apparently critical properties such as the GR law \cite{OFC,Grassberger,Lise} or the  Omori law \cite{Hergarten}, although there still remains controversy concerning whether the model is strictly critical \cite{Lise} or only approximately so \cite{Prado,Boulter}. In this way, the OFC model has been regarded as a typical non-conservative model exhibiting the self-organized criticality (SOC). 

 In the OFC model,  ``stress'' variable $f_i$ ($f_i\geq 0$) is assigned to each site on a square lattice with $L\times L$ sites. Initially,  a random value in the interval [0,1] is assigned to each $f_i$, while $f_i$ is increased with a constant rate uniformly over the lattice until, at a certain site $i$, the $f_i$ value reaches a threshold, $f_c=1$. Then, the site $i$ ``topples'', and a fraction of stress $\alpha f_i$ ($0<\alpha<0.25$) is transmitted to its four nearest neighbors, while $f_i$ itself is reset to zero.  If the stress of some of the neighboring sites $j$ exceeds the threshold, {\it i.e.\/}, $f_j\geq f_c=1$, the site $j$ also topples, distributing a fraction of stress $\alpha f_j$ to its four nearest neighbors. Such a sequence of topplings continues until the stress of all sites on the lattice becomes smaller than the threshold $f_c$. A sequence of toppling events, which is assumed to occur instantaneously, corresponds to one seismic event or an avalanche. After an avalanche, the system goes into an interseismic period where uniform loading of $f$ is resumed, until some of the sites reaches the threshold and the next avalanche starts. 

 The transmission parameter $\alpha$ measures the extent of non-conservation of the model. The system is conservative for $\alpha =0.25$, and is non-conservative for $\alpha <0.25$. A unit of of time is taken to be the time required to load $f$ from zero to unity.

 In the OFC model,  boundary conditions play a crucial role. For example,  SOC state is realized under open boundary conditions, but is not realized under periodic boundary conditions.  Middleton and Tang (MT) observed that the model under open boundary conditions went into a special transient state where events of size 1 (single-site events) repeated periodically with period $1-4\alpha$  \cite{Middleton}. These single-site events occur in turn in a spatially random manner, but after time $1-4\alpha$, the same site topples repeatedly. Although such a periodic state consisting of single-site events is a steady state  under periodic boundary conditions, it is not a steady state under open boundary conditions because of the boundary. Indeed, clusters are formed near the boundary, within which the stress values are more or less uniform, and gradually invades the interior destroying the periodic state. Eventually, such clusters span the entire lattice, giving rise to an SOC-like steady state. MT pointed out that such clusters might be formed via synchronization between the interior site and the boundary site, the latter having a slower effective loading rate due to the boundary. 

 Large-scale synchronization occurring in the steady state of the OFC model was further investigated by Bottani and Delamotte, who observed partial and unstable synchronization of clusters of all length scales, {\it i.e.\/}, synchronized clusters of various sizes exist in the steady state with a large, but finite lifetime \cite{Bottani}. The true nature of spatiotemporal correlations of the OFC model, including its criticality, synchronization and possible periodicity, however, deserves further clarification.
 
 In this situation, we perform in the present letter a quantitative study of spatiotemporal correlations of the OFC model by measuring the {\it local\/} recurrence-time (waiting-time) distribution of avalanches \cite{Mori,Ohmura}. The {\it global\/} recurrence-time distribution of the OFC model was studied, {\it e.g.\/}, in Ref.\cite{Hergarten}, which exhibited a monotonic behavior without any periodic feature. In case of the globally defined recurrence time, the next avalanche to measure recurrence may occur anywhere on the entire lattice. In view of the spatial correlations developed in the steady state of the OFC model, and also in view of the ordinary sense of earthquake recurrence in real life, it might be natural to introduce the recurrence time $T$ and its distribution function $P(T)$ {\it locally\/}. Thus, one may define the local recurrence-time as the time passed until the next avalanche occurs in a vicinity of the preceding avalanche, say, within distance $r$ (in units of lattice spacing) of the triggering site (epicenter) of the preceding event.

 We compute the local recurrence-time distribution $P(T,r)$ in the steady state for several values of $\alpha$, with varying the range parameter $r$. In our simulations, we adopt open boundary conditions, and employ the ``pseudo sequential'' algorithm of Ref.\cite{Pinho}. The lattice size is $L=256$, while lattices $L=128$ and 512 are also studied to check the possible system-size dependence. Discarding initial $2\times 10^9$ avalanches to reach the steady state, measurements are made over the subsequent $10^8$ avalanches.

 The computed $P(T,r)$ is shown in Fig.1(a) for the case of $\alpha =0.2$ on a log-log plot, with varying the parameter $r$ in the range 5$\leq r\leq 80$. We are concerned here with relatively large events, considering the local recurrence time for avalanches whose size is greater than 100, {\it i.e.\/}, $s\geq s_c=100$. The size of an avalanche is measured by the total number of topplings involved in the event. (Note that the same site may topple more than once in a single avalanche.)  When the range parameter $r$ is large enough, the local recurrence-time distribution should reduce to the global recurrence-time distribution. Indeed, $P(T,r)$ for $r=80$ exhibits a monotonic behavior similar to that of the global recurrence-time distribution studied earlier \cite{Hergarten}. For smaller values of $r$, however, significantly different behaviors arise: $P(T,r)$ develops a sharp $\delta$-function-like peak at $T=T^*=0.2$, which tends to grow as $r$ gets smaller. This observation suggests that many, though not all, events tend to repeat with a fixed  time-interval $T=T^*$.  Interestingly, the observed ``period'' $T^*=0.2$ for $\alpha =0.2$ is precisely equal to $1-4\alpha$. For smaller $r$, $P(T,r)$ exhibits sub-peaks at multiples of $T^*=0.2$. This is due to the fact that, when $r$ is small and the neighborhood-range is taken narrow, the ``next'' event occurring at $T=T^*$ after the first one might occur slightly outside this neighborhood, and the ``second-next'' event occurring $T=2T^*$ after the first one counted as the next event in measuring the local recurrence time.

\begin{figure}[ht]
\begin{center}
\includegraphics[scale=0.45]{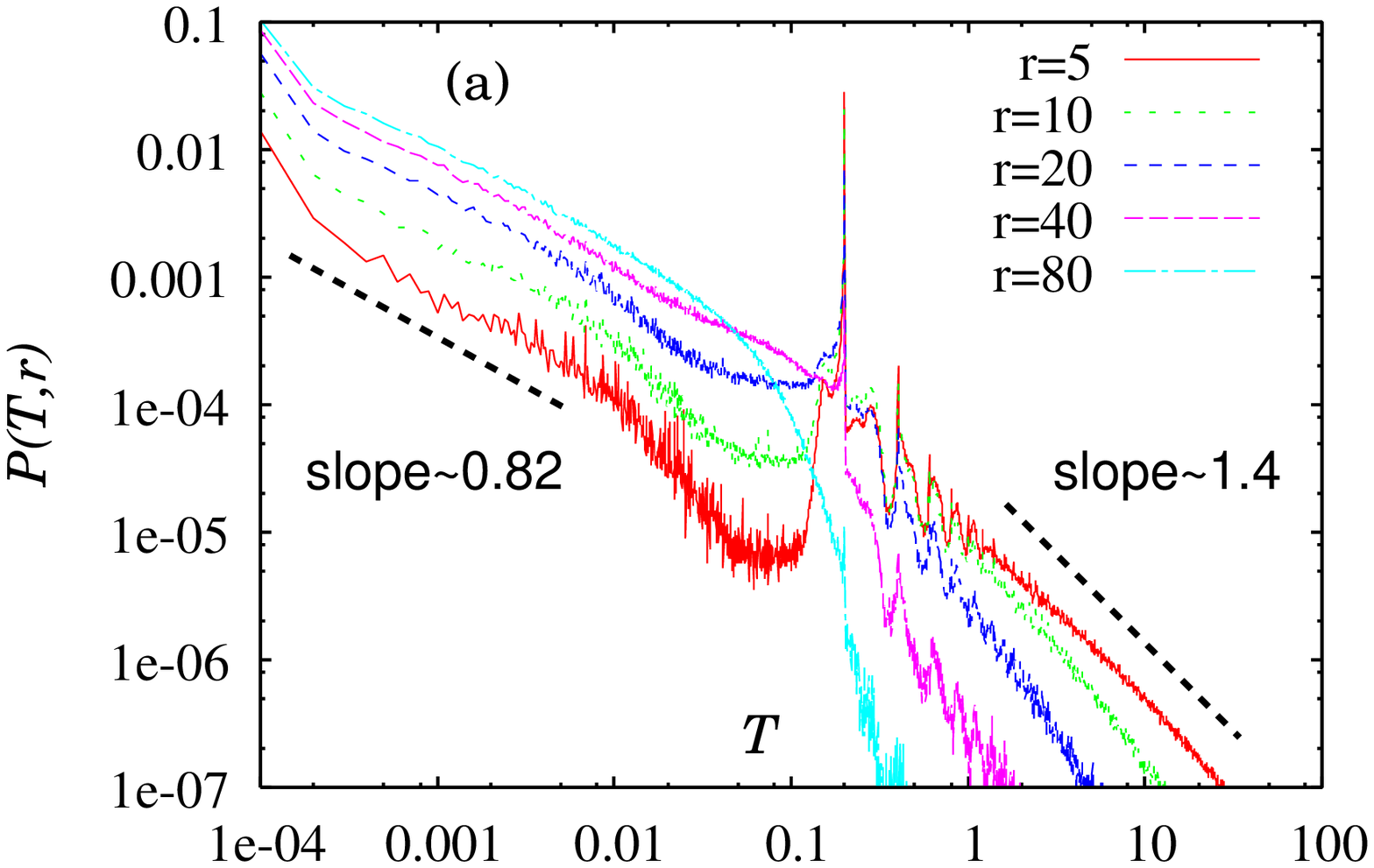}
\includegraphics[scale=0.45]{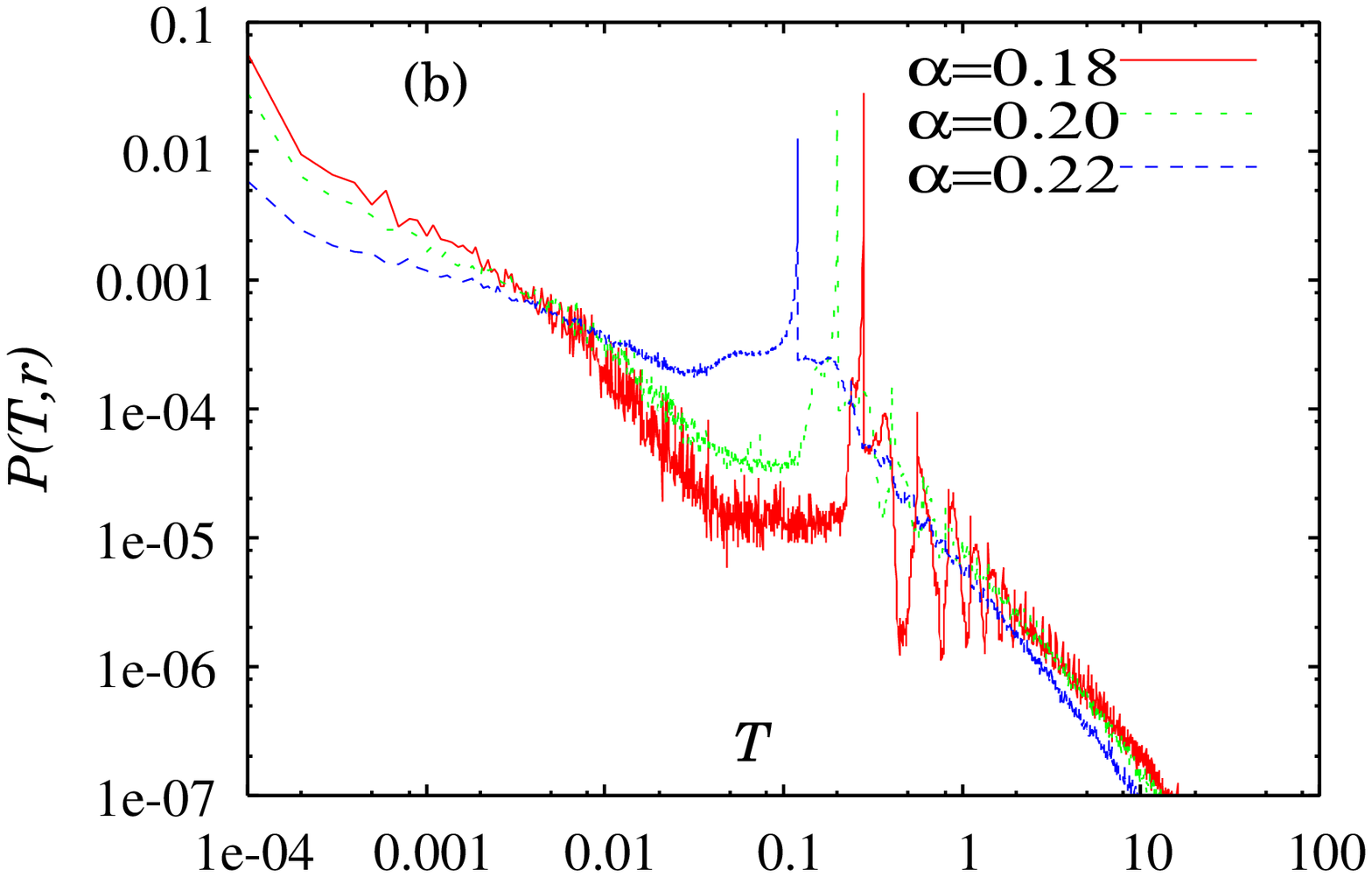}
\includegraphics[scale=0.45]{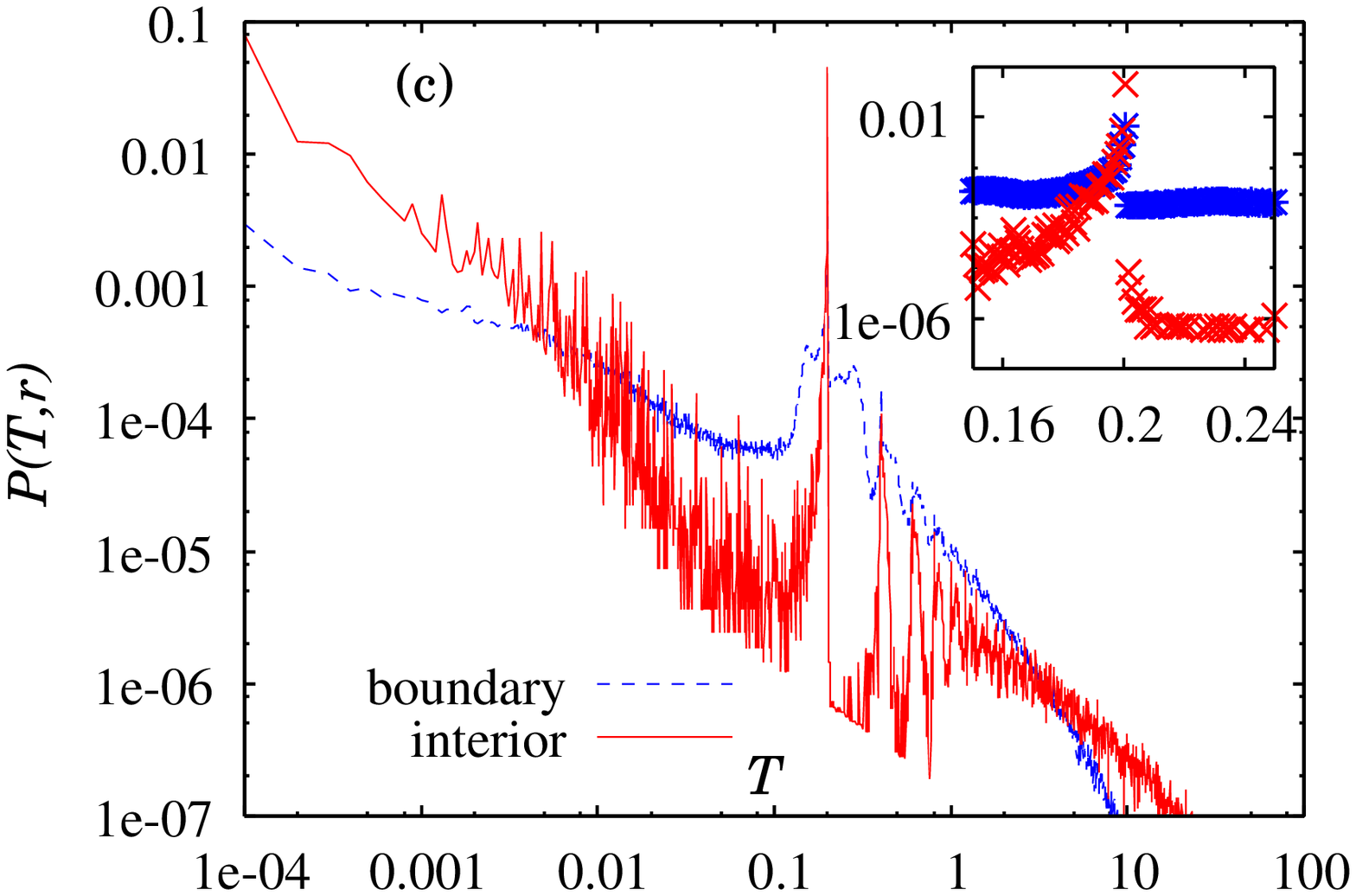}
\end{center}
\caption{
Log-log plots of the local recurrence-time distributions of large avalanches of their size $s\geq s_c=100$; (a) with varying the range parameter $r$ for a fixed $\alpha =0.2$, (b) with varying the transmission parameter $\alpha$ for a fixed $r=10$, and (c) for the interior events whose triggering sites lie in the middle part of the lattice of its size $\frac{2}{5}L\times \frac{2}{5}L$, and for the boundary events whose triggering sites lie in the  boundary region of the width $\frac{1}{5}L$, the parameters being $\alpha=0.2$ and $r=10$. The lattice size is $L=256$.
}
\end{figure}

 It also turns out that, in the shorter and longer time regimes, $T<T^*$ and $T>T^*$, $P(T,r)$ exhibits behaviors close to power laws. In the short-time regime ($T<T^*$), the exponent describing the power law is almost $r$-independent and is equal to $\sim 0.82$, whereas, in the long-time regime ($T>T^*$), it is $r$-dependent and is $\sim 1.4$ for $r=5$, tending to increase with increasing $r$. We have also checked the possible system-size dependence of the result by studying lattices $L=128$ and 512, to observe that the result does not depend on $L$, at least in the time range studied. In this way, the characteristic or periodic feature, {\it i.e.\/}, a sharp peak in $P(T,r)$ at $T=T^*$, and the critical feature, {\it i.e.\/}, power-law behaviors in $P(T,r)$ at $T<T^*$ and at $T>T^*$, coexist in the local recurrence-time distribution.

  In order to see how the observed behavior of $P(T,r)$ depends on the parameter $\alpha$, we also compute $P(T,r)$ for other values of $\alpha$, $\alpha=0.18$ and 0.22, and the results are shown in Fig.1b for the case of $r=10$. As can be seen from the figure, all qualitative features of $P(T,r)$ remain the same in the range of $\alpha$ studied. In particular, the location of the sharp peak turns out to be precisely given by $T^*(\alpha)=1-4\alpha$. The power-law feature is also observed in the short- and long-time regimes for both cases of $\alpha=0.18$ and 0.22. The power-law behavior is particularly eminent in the long-time regime, and the exponent there turns out to be almost $\alpha$-independent (though $r$-dependent). 

 So far, we have considered relative large avalanches of $s_c=100$. However, qualitatively similar behaviors are observed in $P(T,r)$ even when the event-size threshold $s_c$ is varied. In particular, even if one sets $s_c=1$, a sharp peak is still discernible at $T=T^*$, though in a less pronounced manner.

 In order to get further insight into the periodicity versus criticality issue, we next concentrate on the periodic events contributing to a sharp peak of $P(T,r)$ ("peak events"), and investigate the size distribution of these periodic events. More specifically, we record for the $\alpha=0.2$ case the size distribution of avalanches of arbitrary size ($s_c=1$) which occur $T=T^*=0.2$ after the preceding avalanche of arbitrary size in its vicinity of $r=10$. As shown in Fig.2, the size distribution of these peak events exhibit near-critical behavior, not much different from the size distribution obtained for all events without any constraint on its time and position. It means that, even if ones looks at peak events occurring with a fixed ``period'' $T=T^*$, its size distribution exhibits a power-law behavior. Thus,  characteristic feature (in time) and critical feature (in size) also coexist in the OFC model. 

\begin{figure}[ht]
\begin{center}
\includegraphics[scale=0.45]{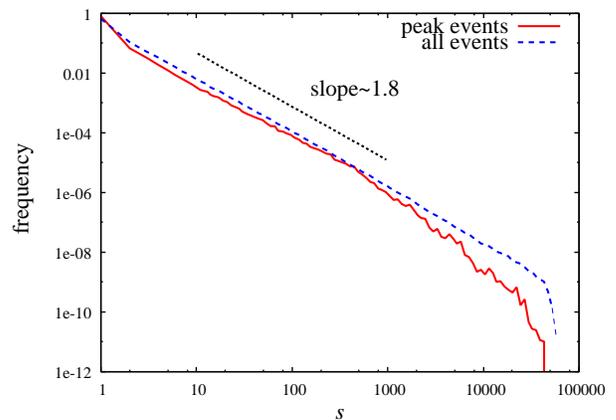}
\end{center}
\caption{
Size distribution of ``periodic events'', {\it i.e.\/}, events of arbitrary size which occur $T=T^*=0.2$ after the preceding event in its vicinity within distance $r=10$. Size distribution without any constraint is also shown for comparison. The lattice size is $L=256$ and the transmission parameter is $\alpha =0.2$.
}
\end{figure}

The peak position of $P(T,r)$ is precisely given by $T^*(\alpha)=1-4\alpha$ irrespective of the values of the range parameter $r$, the size threshold $s_c$ and the system-size $L$. This hidden ``period'' in the steady state coincides with the period of single-site events realized at the transient stage. Of course, periodic recurrence of single-site events is peculiar only to the transient stage, and goes away in the steady state. Furthermore, the periodic single-site events never yields a peak in $P(T,r)$, since successive single-site events does not accompany spatial correlations. Hence, the rhythmic behavior we have observed in $P(T,r)$ in the steady state clearly differs in character from the periodic behavior realized at the transient stage. Yet, it seems natural to suppose that the hidden ``period'' $T^*=1-4\alpha$ in the steady state has  been inherited from the period of single-site events at the transient stage via the formation and invasion process of synchronized clusters.

 It was argued in Refs.\cite{Middleton,Bottani} that synchronization might occur between the interior site, which has a natural period $1-4\alpha$, and the boundary site, which  has a natural period $1-3\alpha$. In the simplest situation,  direct synchronization between two such oscillators yields a period corresponding to the slower one, {\it i.e.\/}, $T=1-3\alpha$. In contrast to this expectation, the peak in $P(T,r)$ turns out to be located precisely at $T=1-4\alpha=T^*(\alpha)$. In Fig.1(c), we show  $P(T,r)$ in the steady state  for the case of $\alpha =0.2$, each measured in the interior and in the boundary of the lattice: In the former, the triggering sites of the two successive avalanches both lie in the inner region of size $\frac{2}{5}L\times \frac{2}{5}L$, while, in the latter, the triggering sites of the two successive avalanches both lie in the boundary region of the width $\frac{1}{5}L$. As can clearly be seen from the figure, the peak is again located precisely at $T=0.2=T^*$ both for the boundary and inner regions. As shown in the inset, the sharp peak at $T=T^*$ is actually not infinitely sharp, with certain finite width. However, the peak itself is always located at $T=T^*$. The detailed mechanism producing such a sharp peak precisely located at $T=T^*$ remains to be understood.

\begin{figure}[ht]
\begin{center}
\includegraphics[scale=0.5]{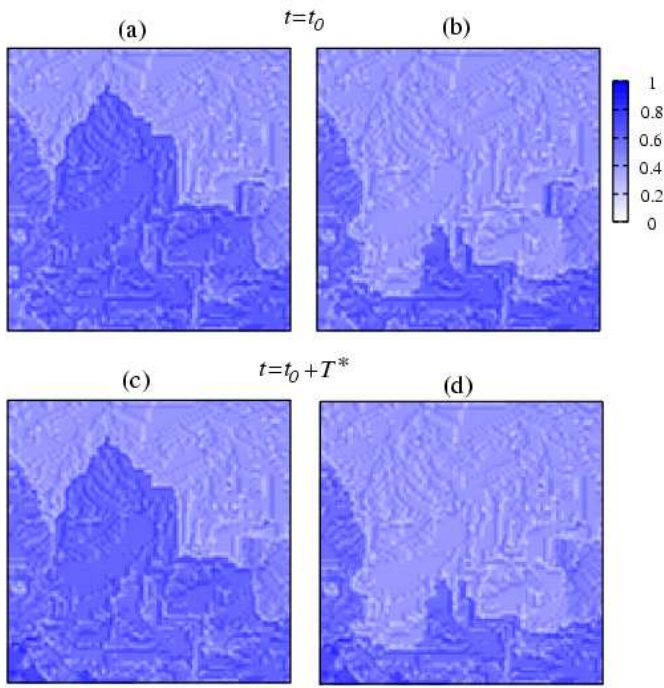}
\end{center}
\caption{
Snapshots of the stress distribution for the case of $\alpha=0.2$; (a) immediately before a large event at time $t=t_0$, (b) immediately after this event, (c) immediately before the following event which occur at time $t=t_0+T^* (T^*=0.2)$, and (d) immediately after this second event. Two events are of size $s=3512$ and $s=3525$ on a $L=256$ lattice. 
Only a part of the lattice is shown in the figure. 
}
\end{figure}

 One may describe the observed phenomena in terms of synchronized clusters.  Synchronized clusters of various sizes exist in the steady state of the OFC model, and governs the steady-state dynamics \cite{Middleton,Bottani}. The sites belonging to the same cluster often topple simultaneously in the same avalanche, {\it i.e.\/}, synchronization. Furthermore, we have observed that some of these synchronized clusters rupture repeatedly with a fixed time-interval $T^*=1-4\alpha$. Such periodic or rhythmic rupture of a given cluster, however, does not repeat permanently.  Indeed, many clusters are left out of the rhythmic recurrence, and rupture more critically with widely-distributed recurrence time, yielding the observed power-law part of $P(T,r)$. 


 In the upper panel of Figs.3, we show for the case of $\alpha=0.2$ typical snapshots of the stress distribution immediately before and after a large event which occur at time $t=t_0$. Discontinuous drop of the stress associated with a rupture of a synchronized cluster is discernible. Then, at time $t=t_0+T^*$, the same cluster (except for a minor difference) ruptures again. In the lower panel of Figs.3, we show snapshots of the stress distribution immediately before and after this subsequent avalanche occurring at $t=t_0+T^*$. In this particular example,  a rhythmic rupture of essentially the same cluster has repeated more than ten times. 

 Finally, we note that the phenomena observed here closely resemble those of the ``asperity'' familiar in seismology \cite{Scholzbook}, in the sense that almost the same spatial region ruptures repeatedly with some period. Although the origin of the asperity is usually ascribed to possible inhomogeneity of the material property of the crust or of the external conditions of that particular region, we stress here that, in the present OFC model, there is no buit-in inhomogeneity in the model parameters nor in the external conditions. The ``asperity'' in the OFC model has been self-generated from the spatially uniform evolution-rule and model parameter. As mentioned, the asperity in the OFC model is not a permanent one: In long terms, its position and shape change. After all, the model is uniform. Nevertheless, recovery of spatial uniformity often takes a long time, and the asperity exists stably over many earthquake recurrences. Although one has to be careful in immediately applying the present result for the OFC model to real earthquakes, it might be instructive to recognize that the observation of asperity-like earthquake recurrence does not immediately mean that the asperity region possesses different material properties nor different external conditions from other regions.

 In summary, we have studied spatiotemporal correlations of the OFC model, and found that its local recurrence-time distribution exhibits a sharp peak at a fixed time $T^*(\alpha)=1-4\alpha$ irrespective of the values of the range parameter, the event-size threshold and the lattice size, together with power-law behaviors at short- and long-time regimes. The size distribution of events repeating with a characteristic time-interval $T^*(\alpha)=1-4\alpha$, however, exhibits a power-law behavior. Hence, the periodic or characteristic feature and the critical feature coexist in the OFC model in an intriguing manner.

\end{document}